\documentclass[12pt]{iopart}
\usepackage{iopams}  
\newcommand{\ensemblemean}[1]{\langle\hskip-2pt\langle{#1}\rangle\hskip-2pt\rangle}

\usepackage{graphicx}
\usepackage{dcolumn}
\usepackage{bm}

\usepackage{graphics}

\begin{document}
\title{Critical behavior of a Ginzburg-Landau model with additive quenched noise}
\author{Niko Komin, Lucas Lacasa, Ra\'{u}l Toral}
\address{IFISC (Instituto de F\'{\i}sica Interdisciplinar y Sistemas Complejos), CSIC-Universitat de les Illes Balears, Campus UIB, 07122, Palma de Mallorca, Spain}

\begin{abstract}
We address a mean-field zero-temperature Ginzburg-Landau, or $\phi^4$, model subjected to quenched additive noise, which has been used recently as a framework for analyzing collective effects induced by diversity. We first make use of a self-consistent theory to calculate the phase diagram of the system, predicting the onset of an order-disorder critical transition at a critical value $\sigma_c$ of the quenched noise intensity $\sigma$, with critical exponents that follow Landau theory of thermal phase transitions. We subsequently perform a numerical integration of the system's dynamical variables in order to compare the analytical results (valid in the thermodynamic limit and associated to the ground state of the global Lyapunov potential) with the stationary state of the (finite size) system. In the region of the parameter space where metastability is absent (and therefore the stationary state coincide with the ground state of the Lyapunov potential), a finite-size scaling analysis of the order parameter fluctuations suggests that the magnetic susceptibility diverges quadratically in the vicinity of the transition, what constitutes a violation of the fluctuation-dissipation relation. We derive an effective Hamiltonian and accordingly argue that its functional form does not allow to straightforwardly relate the order parameter fluctuations to the linear response of the system, at odds with equilibrium theory. In the region of the parameter space where the system is susceptible to have a large number of metastable states (and therefore the stationary state does not necessarily correspond to the ground state of the global Lyapunov potential), we numerically find a phase diagram that strongly depends on the initial conditions of the dynamical variables. Specifically, for symmetrically distributed initial conditions the system evidences  a disorder-order transition for $\sigma_c'<\sigma_c$, yielding a reentrant transition in the full picture. The location of $\sigma_c'$ increases with parameter $a$ and eventually coalesces with $\sigma_c$, yielding in this case the disappearance of both transitions. On the other hand, for positive-definite initial conditions the order-disorder transition is eventually smoothed for large values of $a$, and no critical behavior is found accordingly. At this point we conclude that structural diversity can induce both the creation and annihilation of order in a nontrivial way. 
\end{abstract}
\date{\today}
\maketitle

\section{Introduction}
In statistical mechanics, models describing the effect of impurities or heterogeneities in the behavior of magnetic systems are gathered under the label of spin glasses \cite{young} when the source of heterogeneity affects the local spin interaction (and therefore the interaction term in the Hamiltonian takes into account such disorder). Conversely, the so-called random field models \cite{young} address those systems where the source of heterogeneity only yields an additive heterogeneous term (perturbation) in the Hamiltonian: in this case the effect of disorder is akin to subject the system to a random external perturbation. In both cases, such sources of heterogeneity typically have slower dynamical evolution than the spins (or the dynamical variables), and therefore these sources of randomness are said to be quenched. In the last decades a wealth of literature has addressed the phenomenology behind spin glasses and random field models, including phase diagrams, aging and other dynamical behavior, and comparison with their equilibrium counterparts (see \cite{young, spin1, spin2} and references therein).

In other branches of science the role of disorder in models characterizing the dynamical behavior of multicomponent systems has also been addressed in the last years. Noticeable examples include the effect that a certain amount of heterogeneity in the natural frequencies of Kuramoto oscillators can yield on synchronization \cite{kuramoto1, kuramoto2}, the paradoxical constructive role that disorder can induce in the formation of ordered structures in a plethora of different contexts \cite{buceta2001, tessone2006, tessone2007a, ToralTessoneLopes2007, toral2008a, chen2008, Gosak2009, Ullner2009, Zanette2009, Tessone2009, Postnova09, Perc08}, and the effect that the topology of the underlying network of interactions plays in several types of dynamics \cite{yamir, Chen09, Wu09, Acebron07}, to cite some. All these works address similar generic questions, namely study the effect of structural disorder in the dynamics of multicomponent systems.

In this paper we will address a paradigmatic example within equilibrium statistical mechanics, the Ginzburg-Landau, also called $\phi^4$, model \cite{amit}, in a version subjected to such quenched disorder much in the vein of random field models.  Although the expected role of heterogeneity is that of destroying the ordered state, recent works \cite{tessone2006, ToralTessoneLopes2007} have addressed the positive role of the quenched noise in enhancing the response of this model under the presence of an external periodic driving. In \cite{buceta2001} the authors studied the effects of introducing a quenched multiplicative dichotomous noise, and found that the phase diagram is modified and gives rise to the onset of reentrant phase transitions not present in the quenched noise free model.

Here we address the mean-field version of the model subjected to quenched additive noise in absence of temperature \cite{ToralTessoneLopes2007, Komin}. First, we present an analytical study of the phase diagram by means of a self-consistent theory, both in the non-metastable and metastable regions. The theory predicts an order-disorder transition as a function of the quenched noise intensity $\sigma$, with mean field critical exponents equal to those of  the thermal equilibrium counterpart. We also perform a detailed numerical study of the system for different sizes $N$ in terms of finite-size-scaling theory and determine the scaling exponents. We show that  in the non-metastable region the order parameter fluctuations diverge with an exponent different from the one of the magnetic susceptibility. This indicates a violation of the fluctuation-dissipation relation. In order to justify this finding, we obtain in closed form an expression for the probability density function of the system in terms of an effective Hamiltonian ${\cal H}_{\rm eff}({\bf x})$, and accordingly argue that the fluctuations of the order parameter cannot be straightforwardly related to the linear response of the system. In the region where metastability does take place, results from numerical simulations deviate from the phase diagram found through the self-consistent theory and show a strong dependence on the specific initial conditions. Concretely, we show that for symmetrical initial conditions, the simulations point out the presence of a reentrant phase transition (disorder-order-disorder) with an ordered state whose width varies and eventually disappears in the Ising limit, corresponding to a large valued of a parameter in the Hamiltonian.  This counterintuitive phenomenology supports the fact that disorder or heterogeneity can not only induce dynamical disorder but, on the contrary, can have an ordering role. Conversely, for positive-definite initial conditions the phase transition is smoothed in the same limit, and no critical behavior is found in that case.

The rest of the paper is organized as follows: in section \ref{preliminary} we present the model. In section \ref{metastability} we outline some considerations regarding the presence of metastable states. In section \ref{critical} we derive the mean-field critical exponents associated to the magnetization and magnetic susceptibility. In section \ref{results1} we numerically study the order-disorder transition in the range of parameters where the system lacks metastable states. We provide compelling evidence suggesting that the fluctuation-dissipation relation is not satisfied, and we argue that a possible reason is that the influence of the average external field $h$ on the effective Hamiltonian yielding the probability density function of the system cannot be readily stated as  ${\cal H}_{\rm eff}({\bf x})={\cal H}_0({\bf x})+Nmh$, being $m$ the magnetization, as it happens in equilibrium theory. In section \ref{results2} we numerically explore the system's behavior in the presence of metastable states and discuss the role of the initial conditions in the asymptotic stationary state of the system. We also point out the presence of an disorder-order transition induced by diversity in the metastable situation. In section \ref{conclusions} we summarize our main results.

\section{Additive Ginzburg-Landau model: preliminary considerations}
\label{preliminary}
We consider a set of $N$ real dynamical variables $x_i(t), i=1,\dots,N$ whose evolution is given by a relaxational gradient flow \cite{st00} in a potential $V$:
\begin{eqnarray}
&&\frac{dx_i}{dt}=-\frac{\partial V(\mathbf{x};\mathbf{\eta})}{\partial x_i}\nonumber \,,\\
&&V=\sum_{i=1}^N\left[-\frac{a}{2}x_i^2+\frac{1}{4}x_i^4+\frac{1}{4N}\sum_{j=1}^N(x_j-x_i)^2-\eta_ix_i\right],\label{Lyapu}
\end{eqnarray}
or,
\begin{equation}
\frac{dx_i}{dt}=ax_i - x_i^3 + \frac{1}{N}\sum_{j=1}^N (x_j-x_i) + \eta_i\,. \label{dxi}
\end{equation}
The Lyapunov potential $V(\mathbf{x};\mathbf{\eta})$ depends, besides on the dynamical variables $\mathbf{x}\equiv (x_1,...,x_N)$, on a set of variables $\mathbf{\eta}\equiv (\eta_1,...,\eta_N)$. Most commonly these variables represent white noise of amplitude proportional to the temperature and the model defines a class of thermal phase transitions. In this work, however, we take these variables to represent quenched noise and the problem then belongs to a class of zero temperature random field models. Accordingly, $(\eta_1,\dots,\eta_N)$ are independently drawn from a probability distribution $g(\eta)$ (which typically will  be a Gaussian) of mean $h$ and standard deviation $\sigma$. The model can be thought as describing a set of globally coupled heterogeneous units, being $\sigma$ a measure of the amount of diversity or heterogeneity in the system. As we are interested in this work in the effect of the diversity, $\sigma$ will be taken as a control parameter and we will study the effect that $\sigma$ has on the collective properties of the system.

This model is indeed a discretization of a mean--field version of the well known Ginzburg-Landau Hamiltonian for a scalar field $x(\vec r)$ under the presence of a random external field $\eta(\vec r)$ \cite{young, amit}:
\begin{equation}
{\cal H}
 = \int d\vec r\left(-\frac{a}{2}x^2+\frac{C}{2}|\vec \nabla x|^2+\frac{u}{4}x^4-\eta x\right), 
\end{equation}
where, without loss of generality, we have rescaled variables and time such that $u=1,\,C=1/2$.  This Hamiltonian provides a coarse-grained description of critical phenomena, and its formulation is based on some phenomenological considerations such as locality and symmetries (rotational and translational); that is to say, this latter expression is not calculated from the microscopic physics, but rather can be understood as a coarse-grained description of the magnetization field $x$. By using the Boltzmann weight factor ${\rm e}^{-{\cal H}/T}$, where $T$ is the temperature, this model has been used for instance to describe the paramagnetic-ferromagnetic transition (where the Hamiltonian describes the coarse-grained magnetization field). In the case of a uniform external field, Landau theory elegantly describes a second-order thermal phase transition for this system, with mean-field critical exponents $\beta=1/2$, $\gamma=1$ \cite{young, amit}.   This Hamiltonian also offers a soft-spin description of the Ising model \cite{young}: as a matter of fact, in the limit $a\rightarrow\infty$ one recovers the Ising model (or the Random Field Ising Model (RFIM) in the case of having a random external field). In the last decades the RFIM has been extensively studied (see \cite{young,sethnareview} and references therein), where some specific results include the onset of criticality in terms of a second-order phase transition in zero-temperature induced by the disorder of the random field, with mean-field critical exponents \cite{sethna1, sethna2} as in the thermal counterpart \cite{schneider}. Several other features such as hysteresis, avalanche dynamics, or return point memory effects, to cite a few, have been studied within the RFIM, both in analytical (renormalization-group) and numerical (finite-size scaling) terms \cite{sethnareview, perez}. The properties of the model have also been studied in the context of domain growth dynamics both in the Ising limit\cite{grantgunton1,grangunton2,gf84} or using the full Ginzburg-Landau Hamiltonian \cite{oguzetal} .

\section{On the presence of metastability}
\label{metastability}
From the dynamical point of view, it has already been said that the evolution is relaxational in the Lyapunov potential $V$. Hence, the absolute minimum (or ground state) of $V$ located at ${\mathbf{\bar x}}\equiv(\bar x_1,\dots,\bar x_N)$ must be considered as the global attractor of the dynamics. It is obvious that the value of ${\mathbf{\bar x}}$ will depend on the specific realization of the quenched noise variables $(\eta_1,\dots,\eta_N)$. On the other hand, the solutions of the differential equations (\ref{dxi}) tend to values $x^{st}_i=\lim_{t\to \infty}x_i(t)$ which might or might not coincide with $\bar x_i$. If the potential $V$ has a single minimum, then the dynamics always leads to $\mathbf {\bar x}$, but if there are additional, metastable, minima, then the asymptotic solution $\mathbf{x^{st}}$ depends on the initial condition $\mathbf{x}(t=0)$ as it might get stuck in one of them. The presence and relevance of these metastable minima depends in general (and besides the particular realization of the quenched-noise variables) on the value of the parameter $a$ and the number of variables $N$. 

In order to find the absolute minimum $\mathbf{\bar x}$ one needs to solve the system of $N$ coupled algebraic equations:
\begin{equation}
0=a{\bar x}_i - {\bar x}_i^3 + \frac{1}{N}\sum_{j=1}^N ({\bar x}_j-{\bar x}_i) + \eta_i. \label{xbar}
\end{equation}
The solution is greatly simplified if one introduces the {\sl magnetization} $m$ as 
\begin{equation}
m=\frac{1}{N}\sum_{i=1}^N {\bar x}_i,\label{mag}
\end{equation}
and then writes Eq.(\ref{xbar}) as:
\begin{equation}
m+ \eta_i=(1-a){\bar x}_i + {\bar x}_i^3. \label{xm}
\end{equation}
This equation allows one to find $\bar x_i$ as a function of $m$ and $\eta_i$ (in fact as a function of $m+\eta_i$). The explicit solution, $\bar x_i=\bar x(m+\eta_i)$ can be replaced in the definition of the magnetization to obtain a {\sl self-consistency} equation:
\begin{equation}
m=\frac{1}{N}\sum_{i=1}^N {\bar x}(m+\eta_i).\label{mag:self}
\end{equation}
The problem has been reduced from the simultaneous solution of the $N$ coupled equations (\ref{xbar}), to the solution of a single one (\ref{mag:self}) although, in general, all possible solutions $m^{(1)},m^{(2)}, \dots$ of this equation have to be obtained numerically. For a given solution $m^{(n)}$ one can then find the respective values of ${\bar x}_i^{(n)}$ using the function ${\bar x}_i^{(n)}=\bar x(m^{(n)}+\eta_i)$. In order to analyze the structure of the possible solutions of Eq.(\ref{mag:self}), it is convenient to split the discussion in the cases $a\le1$ and $a>1$.

\subsection{Case $a\le 1$}
\begin{figure}[!htb]
\centering
\includegraphics[width=0.80\textwidth]{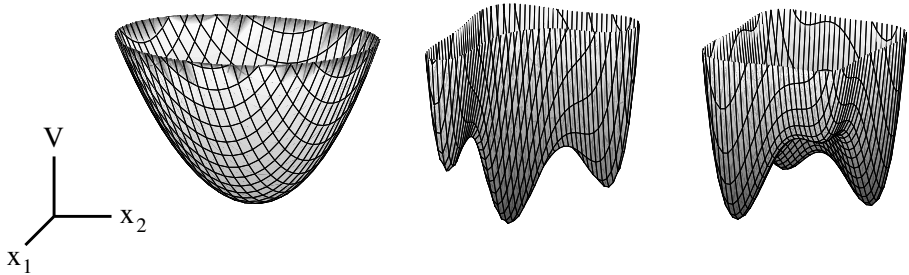}
\caption{Lyapunov potential $V(x_1,x_2)$  as defined in Eq.~(\ref{Lyapu})  for $N=2$,  $\eta_1=-0.48, \eta_2=0.5$ in the cases $a=-1$ (left), $a=0.8$ (center) and $a=2.8$ (right).  While the case $a=-1$ displays a single minimum, in the case $a=0.8$ there are 3 minima (2 metastable) and 2 maxima, whereas for $a=2.8$ there are $4$ minima (3 metastable) and $5$ maxima.} \label{potencial3}
\end{figure}

This is the simplest case. A graphical analysis shows that Eq.(\ref{xm}) has a unique real solution ${\bar x}_i={\bar x}(m+\eta_i)$ (see Appendix). Even in this case, it is possible that Eq.(\ref{mag:self}) has more that one solution for $m$. This is typically the case for small values of $N$.  See an example in Fig.~\ref{potencial3}.

However, as $N$ increases the number of metastable solutions decreases greatly. In fact, it is possible to prove that in the thermodynamic limit, $N\to\infty$, Eq.(\ref{mag:self}) can have only either one or three solutions depending on the values of $a,h,\sigma$. The proof replaces the sum over $N$ by an integral over the probability distribution of the quenched-noise variables:
\begin{equation}
m=\int d\eta\,g(\eta){\bar x}(m+\eta). \label{SC1}
\end{equation}
Let us assume  that the probability distribution $g(\eta)$ has a generic form $g(\eta)=\frac{1}{\sigma}G\left(\frac{\eta-h}{\sigma}\right)$. Henceforth, all numerical results will use the Gaussian distribution $G(z)=\frac{1}{\sqrt{2\pi}}{\rm e}^{-z^2/2}$. A change of variables leads to:
\begin{equation}
m=\int d\xi\,G(\xi){\bar x}(m+h+\sigma \xi)\equiv F_{\sigma}(m+h). \label{SC2}
\end{equation}
As $F_{\sigma}(z)$ is a monotonously increasing function satisfying $F_{\sigma}(0)=0$ and with a sigmoidal shape~\cite{fsigma}, there will be only one solution for $m$ for all values of $h$ if the derivative satisfies $F_{\sigma}'(0)\le 1$. On the other hand, for  $F_{\sigma}'(0)> 1$ there will be either one or three solutions depending on the value of $h$. This analysis mimics that of the Weiss mean-field theory \cite{stanley} and allows one to compute the magnetization $m(h;a,\sigma)$ as a function of the mean value of the disorder $h$ and the parameters $a,\sigma$. It displays usual critical phenomena and hysteresis. The critical point is defined by the condition $F_{\sigma}'(0)=1$ and can be achieved by varying $a$ or $\sigma$. It is possible to show that $F_{\sigma=0}'(0)=1/(1-a)$ and, since $F_{\sigma}'(0)$ is a decreasing function of $\sigma$, the condition $F_{\sigma}'(0)=1$ can never be achieved for $a<0$.  This was a priori obvious since in that case the Lyapunov potential in the absence of quenched noise has the global minimum at $x_i=0,\forall i$, already a disordered state. Some numerical values (for the Gaussian distribution) for the location of the critical diversity $\sigma_c$ as a function of $a$ are: $(a=0.1,\sigma_c=0.19616),\,(a=0.5,\sigma_c=0.50041),\,(a=2/3,\sigma_c=0.595233)$. In the case $a=1$, the Cardano formula simplifies to $\bar x=(m+h)^{1/3}$ and it is possible to perform analytically the integrals (again for a Gaussian distribution for the quenched-noise variables) with the result \cite{ToralTessoneLopes2007} $(a=1,\sigma_c=\left[\frac{\Gamma(1/6)}{2^{1/3}3\pi^{1/2}}\right]^{3/2}=0.7573428\dots)$.

\subsection{Case $a>1$}

The problem in this case is that the cubic equation (\ref{xm}) can have either one or three real solutions depending on whether the discriminant
$\Delta_i=27(m+\eta_i)^2+4(1-a)^3$ is, respectively, positive or negative. Besides, as before,  several values of $m$ can satisfy the self-consistency Eq.(\ref{mag:self}). When there are three solutions for $\bar x_i$, ($\Delta_i<0$, this requires $a>1$) it is not clear a priori which one to chose in order to substitute in the self-consistency relation (\ref{mag:self}). A possibility is to compute the Lyapunov potential $V$ for each of the possible solutions. However, since the maximum number of solutions can be as large as $3^N$, this is not possible to carry out in practice for large $N$. The answer arises when one realizes that  the dynamical equation for $x_i(t)$ can be written also as relaxational in a local potential $v_i(x_i,m)$:
\begin{eqnarray}
&&\frac{dx_i}{dt}=-\frac{\partial v_i(x_i,m)}{\partial x_i},\nonumber\\
&&v_i=\frac{1-a}{2} x_i^2 + \frac14 x_i^4 - (m+\eta_i)x_i+\frac{m^2}{2}.\label{local}
\end{eqnarray}
The solutions $\bar x(m+\eta_i)$ are nothing but the extrema of this local potential. Now we notice that the Lyapunov potential can be written as sum of the local potentials:
 \begin{equation}
V(x_1,\dots,x_N)=\sum_{i=1}^Nv_i(x_i,m).
\end{equation}
Therefore, the absolute minimum of $V$ is achieved by choosing in each case the solution $\bar x(m+\eta_i)$ that minimizes the local potential  $v_i(x_i,m)$. Explicit expressions for the function $\bar x$ are obtained using Cardano's formula and are given in the Appendix.

The process to find the absolute minimum $\bf \bar x$ of the Lyapunov potential proceeds, as before, by finding first $m$ after solving numerically the self-consistency equation (\ref{mag:self}), but using the correct function $\bar x(m+\eta)$. Similarly, the integral equation (\ref{SC1}) can be used to find the magnetization $m(h;a,\sigma)$ in the thermodynamic limit. The phenomenology of the solutions is similar to what was found in the case $a\le1$ and will not be repeated here.

An important difference, however, with the case $a\le 1$ is that now the Lyapunov potential displays a large number of metastable minima for all values of $N$ and, consequently, also in the thermodynamic limit (a recent study for the metastable states of the zero-temperature RFIM has been carried on in \cite{perez2, perez3}).  Therefore, starting from arbitrary initial conditions, the asymptotic solution of the evolution equations $x_i^{st}$ will in general differ from the values $\bar x_i$ of the absolute minimum. It will be shown that new phase transitions occur when one looks at the magnetization values that derive from the stationary solution.

\section{Critical behavior}
\label{critical}
We have seen that this mean-field model displays a second order phase transition between an ordered state ($|m|>0$) and a disordered state ($m=0$)  at a critical value of the diversity $\sigma_c$.  In order to derive the critical exponents of such transition, we consider the self-consistency Eq.~(\ref{SC2}) and expand $F_{\sigma}(m+h)$ in a Taylor series. Since $\bar x(-m-h)=-\bar x(m+h)$ (see Appendix) and assuming that the distribution of noises is symmetric with respect to the mean value, $G(-\xi)=G(\xi)$, the function $F_{\sigma}$ is antisymmetric $F_{\sigma}(-m-h)=-F_{\sigma}(m+h)$ and we get:
\begin{equation}
m=a_1(\sigma) (m+h) + a_3(\sigma)(m+h)^3+\dots
\label{SC3}
\end{equation}
with $a_k(\sigma)=F_{\sigma}^{(k)}(0)/k!$. Hence, the magnetization at $h=0$ is:
\begin{equation}
\label{cases}
|m|=\cases{0&for $\sigma>\sigma_c,$\\
\sqrt{\frac{1-a_1(\sigma)}{a_3(\sigma)}} &  for $\sigma<\sigma_c.$}
\label{spontaneousm}
\end{equation}
As $F_{\sigma}'(0)-1$ changes sign at $\sigma=\sigma_c$, we can expand $a_1(\sigma)=1+\alpha_1(\sigma_c-\sigma)+\dots$.
Accordingly, close to the transition the spontaneous magnetization behaves as $|m|\sim (\sigma_c-\sigma)^{\beta}$,  with a critical exponent $\beta=1/2$, as in Landau's treatment of the thermal phase transition.

To compute the critical behavior of the susceptibility $\displaystyle \chi_h\equiv\left.\frac{\partial m}{\partial h}\right|_{h=0}$, we take the derivative of both sides of Eq.(\ref{SC3}) and set $h=0$. This leads to $\displaystyle \chi_h=\frac{a_1(\sigma)+3a_3(\sigma)m^2}{1-a_1(\sigma)-3a_3(\sigma)m^2}$. Replacing Eq.(\ref{spontaneousm}) and $a_1(\sigma)=1+\alpha_1(\sigma_c-\sigma)+\dots$ we find the critical behavior:
\begin{equation}
\chi_h=A_{\pm}\left|\sigma-\sigma_c\right|^{-1}
\label{chi}
\end{equation}
with critical amplitudes $A_{-}=1/(2\alpha_1)$ for $\sigma<\sigma_c$ and $A_{+}=1/\alpha_1$ for $\sigma>\sigma_c$.  Therefore the susceptibility critical exponent  is $\gamma=1$, the same, not surprisingly, than in Landau's  theory.

\section{Numerical results for $a\le1$: violation of the fluctuation-dissipation relation}
\label{results1}
\begin{figure}
\centering
\includegraphics[width=0.8\textwidth]{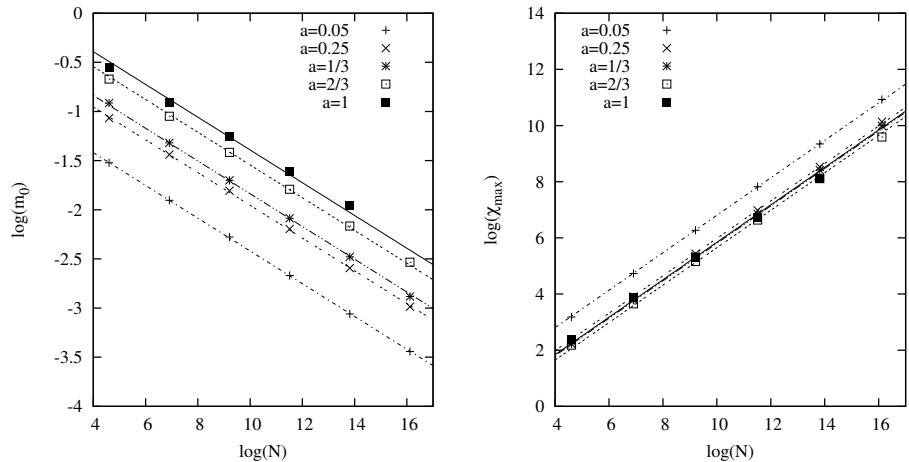}
\caption{Log-log plots of the order parameter $m_0$ (left panel) and the fluctuations $\chi$ (right panel)  as a function of system  size $N$ for different values of $a$, at the corresponding critical point $\sigma_c(a)$.  In all cases we find a good fit to a power-law behavior:  $m_0\sim N^{-c}$ and $\chi \sim N^{b}$ with $c=0.16\pm0.01$ and $b=0.66\pm0.02$.} \label{scaling}
\end{figure}

In this section we present the results coming from the numerical integrations~\cite{numericaldetails} of the dynamical equations (\ref{dxi}) in the case $a\le 1$. The objective is twofold. First, by comparing with the analytical results valid in the thermodynamic limit, we want to check the importance of the metastable states that appear for finite $N$. Second, we will use the theory of finite-size scaling in order to determine the exponents of the transition. We will show that there is a violation of the fluctuation-dissipation relation in the sense that the magnetic susceptibility can not be computed as the ensemble fluctuations of the magnetization. By ensemble average $\ensemblemean{\cdots}$ we mean an average with respect to realizations of the random quenched-noise variables as well as with respect to the initial conditions ${\bf x}(t=0)$. However, for the range of values of system size $N$ employed in the simulations, $N\ge 10^3$, there was hardly any dependence on the initial condition for a given realization of the random variables. This shows that metastable states either do not exist or it is rare to get trapped in them for this range of values of $a$ and $N$.
\begin{figure}
\centering
\includegraphics[width=0.8\textwidth]{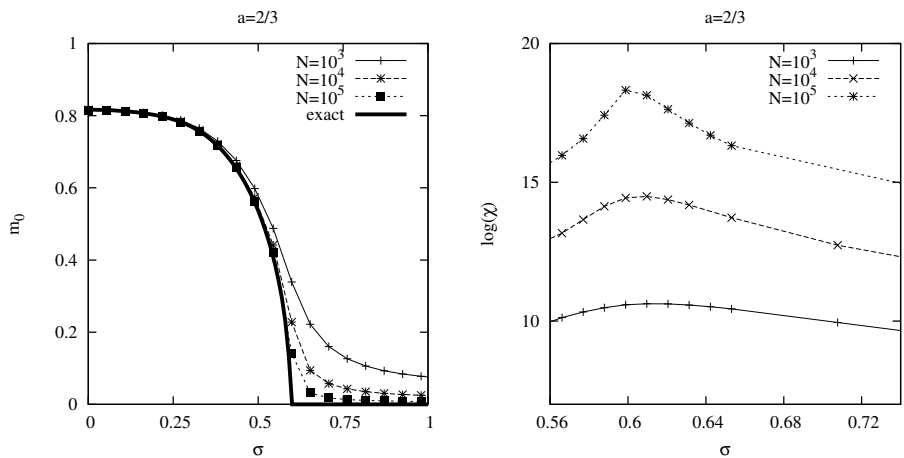}
\caption{\textit{Left panel:} Order parameter $m_0$ as a function of the diversity $\sigma$ for $a=2/3$. The symbols correspond to the numerical integration of the dynamical equations (\ref{dxi}) for different system sizes $N$ and a Gaussian distribution (zero mean, standard deviation $\sigma$) of the quenched noises. The solid line is the magnetization $m$ obtained by solving the self-consistency Eq.(\ref{SC2}) for $h=0$. \textit{Right panel:}
Order parameter fluctuations, $\chi$, as a function of the diversity $\sigma$, for the same system sizes as the left panel (the vertical axis is in logscale for presentation purposes).}\label{figm1}
\end{figure}
In the left panel of Fig.~\ref{figm1}  we plot the order parameter $m_0$ as a function of the diversity $\sigma$ for the value $a=2/3$. As usual \cite{Montecarlo}, the order parameter is defined as the ensemble average of the absolute value of the magnetization $m_0=\ensemblemean{|m|}$ computed from the stationary values as  $m=\frac{1}{N}\sum_{i=1}^Nx_i^{st}$. As predicted by the self-consistent treatment explained in previous sections, there is a phase transition from an ordered  (ferromagnetic-like, $m_0>0$) to a disordered (paramagnetic-like, $m_0=0$) phase as a function of $\sigma$. The transition is smeared out by finite-size effects, but it approaches the solution of the thermodynamic limit and the transition point $\sigma_c$ as the system size $N$ increases. In the right panel of this figure we plot the normalized fluctuations of the order parameter, defined as $\chi\equiv \frac{N}{\sigma^2}\left[\ensemblemean{m^2}- \ensemblemean{|m|}^2\right]$ as a function of the diversity $\sigma$. These fluctuations have a maximum in the neighborhood of $\sigma_c$ and, as shown in the right panel of Fig.~\ref{scaling}, they increase with increasing $N$ as $\chi(\sigma_c)\sim N^{b}$ with $b\approx2/3$ for different values of the parameter $a$, and hence diverge in the thermodynamic limit. As shown in the left panel of the same figure, the order parameter at the critical point decreases as $m_0(\sigma_c)\sim N^{-c}$ with $c\approx 1/6$ and tends to zero in the thermodynamic limit.

\begin{figure}
\centering
\includegraphics[width=0.8\textwidth]{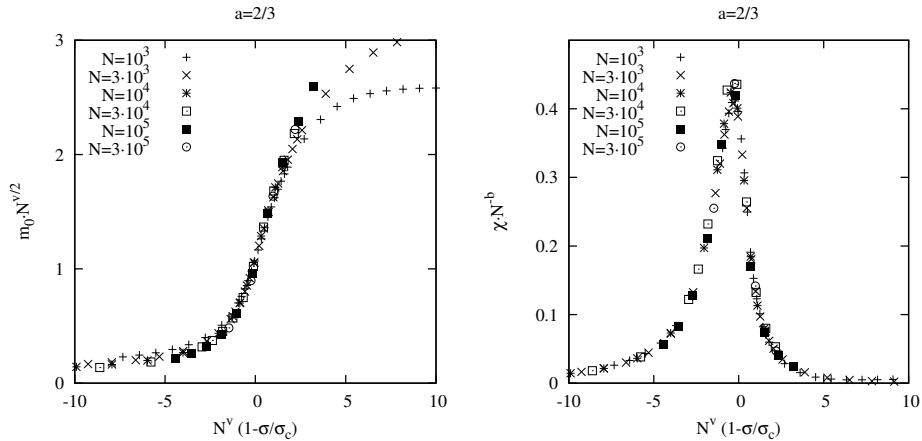}
\caption{Data collapse of the order parameter $m_0$ (left panel) and the fluctuations $\chi$ (right panel) according to the finite-size scaling relations $m_0(\sigma,N)=N^{-v/2}f_m(N^v(1-\sigma/\sigma_c))$ and $\chi(\sigma,N)=N^b f_\chi(N^v(1-\sigma/\sigma_c))$ using $v=1/3$, $b=2/3$. The goodness of the collapse is an evidence supporting the validity of the scaling relations.}\label{scaledfig} 
\end{figure}

Data for a range of values around the critical region can be collapsed through standard finite-size analysis \cite{cardy, deutsch} according to the
scaling laws: $m_0(\sigma,N)=N^{-c}f_m(N^v(1-\sigma/\sigma_c))$ and $\chi(\sigma,N)=N^b f_\chi(N^v(1-\sigma/\sigma_c))$ with appropriate scaling functions $f_m$ and $f_\chi$. A good fit, see Fig.~\ref{scaledfig}, is obtained with $v=2c\approx1/3$. Note that this scaling form implies that in the infinite-size limit $m_0(\sigma)\sim |\sigma-\sigma_c|^{\beta}$ and $\chi(\sigma)\sim |\sigma-\sigma_c|^{-\gamma}$, with critical exponents $\beta=c/v= 1/2$ and $\gamma=b/v= 2$. We have also performed a finite-size scaling of the fluctuations of the stationary value of the energy (global potential) at the critical disorder $\sigma_c(a=2/3)=0.595233$, according to which one finds a value for the critical exponent of those fluctuations $\alpha\approx0$, the same than the (thermal) mean-field result for the specific heat (data not shown). 

While the result of the previous section proved that the susceptibility $\chi_h$ has a critical exponent $\gamma=1$, the numerical simulations suggest that the fluctuations $\chi$ diverge close to the critical point as a power law with a different exponent $\gamma=2$. This seems to constitute a violation of the fluctuation-dissipation relation. Since we have restricted this analysis to the range $a\le 1$, this violation does not seem to be related to typical situations of metastability, absence of time translation symmetry or aging \cite{young, spin1, spin2}. Furthermore, the hyperscaling relation $2\beta+\gamma=d_c \nu$, that holds in the mean-field regime or for $d\geq d_c$, is satisfied using $\gamma=2$ as it is known \cite{grinstein} that the upper critical dimension is $d_c=6$ and $\nu=1/2$.

To explain this discrepancy, we note that the fluctuation-dissipation relation is obtained typically for a system in the canonical ensemble at temperature $T$ and whose probability density function (pdf) is $f_{\bf{x}}={\cal Z}^{-1}\exp(-{\cal H}/T)$, with a partition function ${\cal Z}=\int d{\bf x}\,\exp(-{\cal H}({\bf x})/T)$, being ${\cal H}$ the Hamiltonian of the system. If the Hamiltonian contains a magnetic interaction ${\cal H}({\bf x})={\cal H}_0({\bf x})+Nmh$, one can prove the fluctuation-dissipation relation between the magnetic susceptibility $\chi_h$ and the fluctuations of the magnetization $\langle m\rangle$:

\begin{equation}
\chi_h\equiv\frac{\partial \langle m \rangle}{\partial h}\bigg|_{h=0}=\frac{N}{T}\left[\langle m^2\rangle - \langle m\rangle^2\right], \label{fdt}
\end{equation}
where $\langle \cdots\rangle$ denotes an average with respect to the probability distribution $f_{\bf x}({\bf x})$.

In our case, there are two averages: with respect to initial conditions and with respect to realizations of the random variables ${\mathbf \eta}$. We have already argued that for $a\le1$ and large values of $N$, the results are largely independent of initial conditions, so all that contributes to the ensemble average $\ensemblemean{\cdots}$ are the noise variables. As there is a one to one correspondence between the stationary values $\bar{\bf x}$ and $\bf{\eta}$ we can write the pdf of $\bar{\bf x}$ in terms of the pdf of $\bf{\eta}$:
\begin{equation}
f_{\bf{x}}({\bar x}_1,\cdots,{\bar x}_N)=f_{\bf{\eta}}(\eta_1,\cdots, \eta_N)\left|J\right|.
\label{pdfx}
\end{equation}
If we take the $\eta_i$'s to be independently distributed Gaussian variables, we have
\begin{equation}
f_{\bf{\eta}}(\eta_1,\cdots, \eta_N)=\prod_{i=1}^N\left[\frac{1}{\sigma\sqrt{2\pi}} \exp(-(\eta_i-h)^2/2\sigma^2)\right].
\label{pdfeta}
\end{equation}
As Eq.(\ref{xm}) implies
\begin{equation}
\eta_i= (1-a) {\bar x}_i+{\bar x}_i^{3}-\frac{1}{N}\sum_{j=1}^N {\bar x}_j,
\label{etax}
\end{equation}
it is possible to compute the determinant of the Jacobian matrix $J_{ij}=\frac{\partial \eta_i}{\partial \bar{x}_j}$:
\begin{equation}
\left|J\right|=\left(1-\frac{1}{N}\sum_{j=1}^N\frac{1}{3{\bar x}_j^{2}+1-a}\right)\prod_{i=1}^N \left[3{\bar x}_i^{2}+1-a\right].
\label{jacobian}
\end{equation}
Replacing Eqs.(\ref{pdfeta}-\ref{jacobian}) in Eq.(\ref{pdfx}), one can write the pdf of $\bar{\bf x}$ as the exponential of an effective Hamiltonian $f_{\bar{\bf x}}(\bar{\bf x})={\cal Z}^{-1}\exp(-{\cal H}_{\rm eff})$, with:
\begin{eqnarray}
{\cal H}_{\rm eff}(\bar {\bf x})&=&-\ln\left(1-\frac{1}{N}\sum_{j=1}^N\frac{1}{3{\bar x}_i^{2}+1-a}\right)+\nonumber \\  &&\sum_{i=1}^N\left[\frac{ [(1-a) {\bar x}_i+{\bar x}_i^{3}-m-h]^2}{2\sigma^2}-\ln\left(3{\bar x}_i^{2}+1-a\right)\right]
\end{eqnarray}
However, as it can not be splitted in the form ${\cal H}_{\rm eff}={\cal H}_{0}+Nhm$, it is not possible (at least in a trivial manner) to relate the susceptibility to the fluctuations of the order parameter.

\section{Numerical results for $a>1$: dependence on the initial conditions}
\label{results2}

In the case $a>1$ the presence of metastable states is relevant as the dynamics usually gets trapped in one of them. Therefore, in general, the asymptotic values ${\bf x}^{\rm st}$ depend on initial conditions and the absolute minimum of the potential $V$ might not be reached. Accordingly, deviations from the self-consistent theory are expected to appear. In this section we will study this case and show that a new phenomenology can appear depending on the particular value of $a$ and the distribution of the initial condition ${\bf x}(t=0)$. For the sake of concreteness, we have focused on two types of initial conditions: symmetrical and positive-definite.
\begin{figure}
\centering
\includegraphics[width=0.4\textwidth]{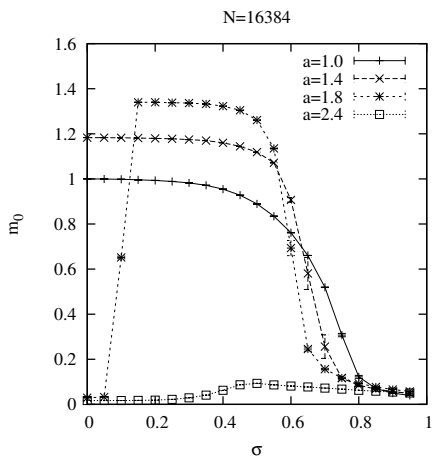}
\caption{Numerical results of the average magnetization as a function of diversity $\sigma$, for a system of $N=16384$ coupled variables for different values of $a\ge 1$ (for the numerical integration of Eq.\ref{dxi}, initial conditions are drawn from a symmetrical uniform distribution $U[-\delta,+\delta]$).
Note that depending on the specific value of parameter $a$, three different behaviors take place: ($I$) an order-disorder transition at $\sigma_c$ for $a=1,1.4$, ($II$) a reentrant phase transition formed by a disorder-order transition at $\sigma_c'$ coupled to an order-disorder one at $\sigma_c$ for intermediate values of $a=1.8$, and ($III$) the absence of any transition to an ordered state for the larger value $a=2.4$.}\label{todas}
\end{figure}
\subsection{Symmetrical initial conditions}

\begin{figure}
\centering
\includegraphics[width=0.8\textwidth, angle=0]{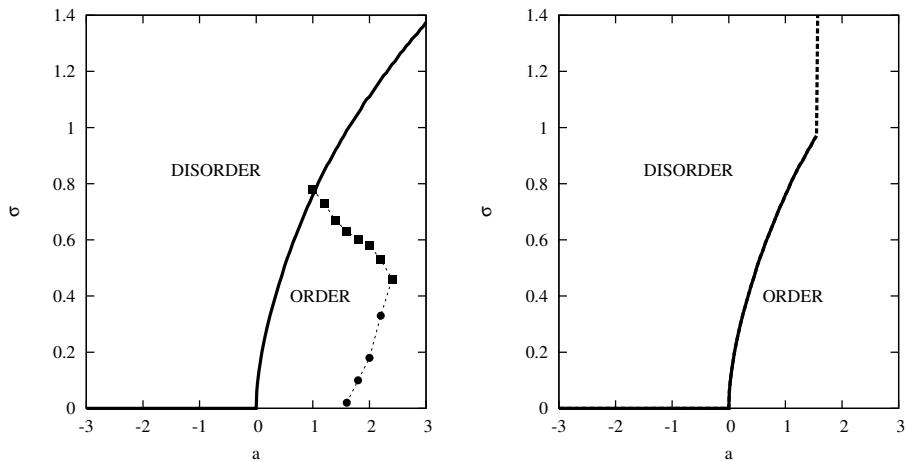}
\caption{\textit{Left panel:} Phase diagram of the system, where the symbols correspond to the values of critical points $\sigma_c$ (associated to the order-disorder transition) and $\sigma_c'$ (associated to the disorder-order transition)
as a function of $a$, for a system of $N=16384$ (derived numerically integrating Eq.\ref{dxi} with initial conditions drawn for a symmetrical uniform distribution $[-\delta,+\delta]$). In the region $a>0$ the system evidences an order-disorder
phase transition at $\sigma_c$, the location of this transition increasing with $a$. The values of $\sigma_c$ (in the thermodynamic limit) can be derived from the self-consistent theory as those satisfying $F'_{\sigma_c}(0)=1$, and are represented by the solid line. In the region $a>1$ the system presents metastable states even in the thermodynamic limit and the solid line refers to the location of phase transition derived from the analysis of the ground state of the Lyapunov potential. At odds with the self-consistent theory, we numerically find for intermediate values of $a$ the coexistence of two phase transitions (reentrant transition), where the location of both critical points converge for increasing values of $a$ until coalescence. At this point the ordered state is completely destroyed for all values of $\sigma$. \textit{Right panel:} Same diagram as for the right panel, when the numerical integration of Eq.\ref{dxi} is performed with initial conditions drawn for a uniform distribution in the positive-definite interval $[0,2\delta]$. In this case, the phase transitions disappear for $a\gtrsim1.4$ as in this case the order parameter $m_0$ tends to zero smoothly with $\sigma$, see right panel of Fig\ref{mfigu2}.}\label{sigC} 
\end{figure}

\begin{figure}
\centering
\includegraphics[width=0.8\textwidth]{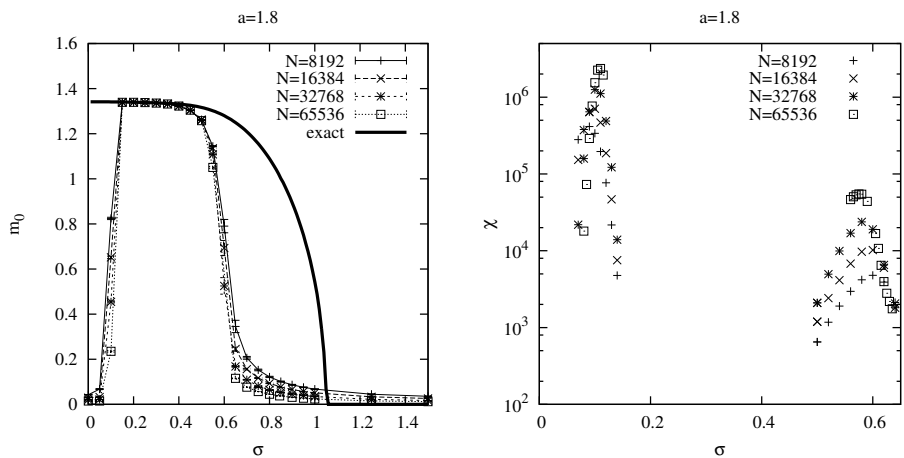}
\caption{\textit{Left panel:} Numerical results of the order parameter as a function of $\sigma$, for different system's size and $a=1.8$, where a reentrant phase transition
takes place (for the numerical integration of Eq.\ref{dxi}, initial conditions are drawn from a symmetrical uniform distribution $U[-\delta,+\delta]$). Exact results from the self-consistent theory are represented in the solid line. The deviations from the theory are related to the fact that the system does not reach the ground state of the Lyapunov potential as it gets trapped in metastable states. \textit{Right panel:} Fluctuations of the order parameter  as a function of $\sigma$ for the same system as the right panel. Fluctuations have
a peaked maximum that scales with system's size close to both transition points.}\label{mfigu} 
\end{figure}

The initial values $x_i(t=0), i=1,\dots,N$, are independently drawn from a uniform distribution in the interval $[-\delta,+\delta]$, for a given value of $\delta$.
In Fig.~\ref{todas} we plot the average magnetization $m_0=\ensemblemean{|m|}$ as a function of  diversity $\sigma$ for different values of $a$ and system size $N=16384$ for $\delta=2.5$. The data have been averaged over $10^2$ initial conditions for ${\bf x }(t=0)$ and then over $10^2$ realizations of the quenched noise variables ($10^4$ averages in total). At variance with the case $a\le 1$ (which is also shown in the figure for comparison) we find three possible scenarios: (i) for $a\gtrsim 1$ (weak metastable regime, $a=1.4$ in the figure) one observes the same phenomenology as for $a\le1$: an order-disorder transition at a critical value $\sigma_c(a)$.  (ii) For larger values of $a$, the former transition is still present at $\sigma_c$, but a new transition (from a disordered state $m_0=0$ to an ordered one $m_0>0$ as $\sigma$ increases) is found at $\sigma_c'<\sigma_c$, see the curve corresponding to $a=1.8$ in the figure. In this case, we find the counterintuitive result that a certain level of diversity in the quenched noise is needed to induce order at $\sigma=\sigma_c'$, whereas a large level of diversity destroys again the ordered state (reentrant phase transition). (iii) Finally, for increasing $a$, $\sigma_c'$  increases and $\sigma_c$ decreases, eventually coalescing for $a>a_c\approx 2.4$, where the ordered state disappears. Thus, for large values of $a$, the system does not evidence any transition and the stationary phase is always the disordered one. We point out that in the curve for $a=2.4$,  the magnetization is not exactly zero for intermediate values of the diversity due to a  finite-size effect:  $m_0$ decreases and approaches zero for all values of $\sigma$ as the system size increases, something that does not occur in cases (i) and (ii). All these features are illustrated in the phase diagram plotted in the left panel of  Fig.~\ref{sigC}: (i) For $1<a\lesssim1.6$ the usual order-disorder transition appears, although the value of $\sigma_c$ is smaller that the one derived from the analysis based upon the structure of the global attractor $\bar{\bf x}$. (ii)  For $1.6\lesssim a \lesssim 2.4$ there is a new transition from a disordered to an ordered state at a value $\sigma_c'<\sigma_c$. (iii) Finally, for $a\gtrsim 2.4$ the only phase encountered is the disordered one.

In order to characterize the transitions that occur in region (ii), we have run extensive simulations for different system sizes in the case $a=1.8$. The order parameter $m_0$ is displayed in the left pane of Fig.~\ref{mfigu}. By looking at the difference with the magnetization curve derived from the theoretical analysis, it is clear from this figure that the system is not able to reach the absolute minimum neither for small or large diversity $\sigma$. We observe at both transitions the same qualitative dependence with system size that was discussed in the case $a\le 1$. As we don't have now a theoretical prediction for $\sigma_c'$ or $\sigma_c$ the numerical analysis of the data is much less conclusive. Pseudo-critical points $\sigma_c(N)$ and $\sigma_c'(N)$ can be defined as the location of the maximum of the fluctuations $\chi$ of the order parameter, see the right panel of Fig.~\ref{mfigu}. The fluctuations scale roughly as $\chi(\sigma_c'(N))\sim N^{b'}$ and $\chi(\sigma_c(N))\sim N^{b}$ with $b'\approx b\approx 0.9$. However, it is difficult to obtain reasonably good quality fits of the data to the standard finite-size-scaling relations used in the case $a<1$. Furthermore, the data show a dependence on $\delta$ (data not shown) such that $\sigma_c$ and $\sigma_c'$ adopt different values for small $\delta$ but saturate for $\delta \gtrsim 2.5$.

Summing up:  if the initial conditions are distributed in a symmetrical interval, the order region is much reduced with respect to the predictions based upon the structure of the ground state. There is a region in parameter space where the system undergoes what appear to be well defined phase transitions, from disorder to order and back to disorder at $\sigma_c'$ and $\sigma_c$, respectively. The order-disorder transition ($\sigma_c$) is related to the one found in the regime $a<1$, while the disorder-order transition (at $\sigma_c'<\sigma_c$) is a new behavior whose nature is genuinely metastable. For $a\gtrsim2.4$ the system is never in the ordered state.

\subsection{Positive-definite initial conditions}

\begin{figure}
\centering
\includegraphics[width=0.8\textwidth]{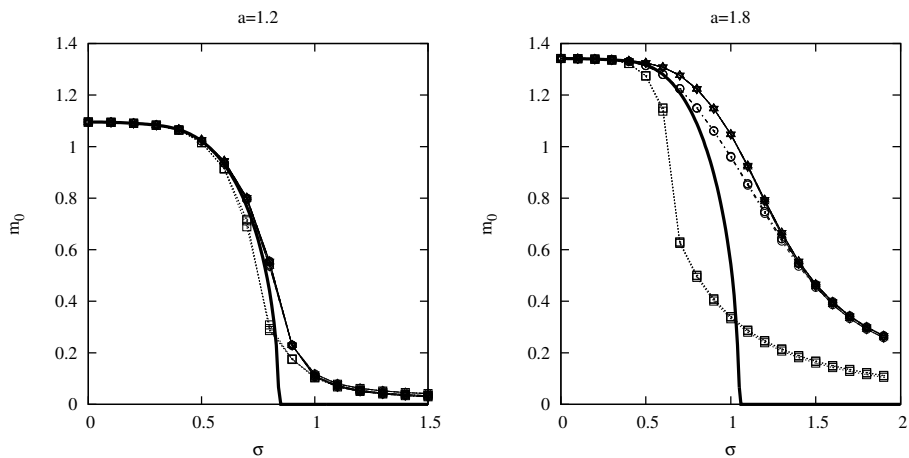}
\caption{Numerical results of the order parameter as a function of $\sigma$, for different system's size $N=4096, 8192, 16384$, and a value $a=1.2$ (left panel) and $a=1.8$ (right panel). For the numerical integration of Eq.\ref{dxi}, initial conditions are drawn from a positive-definite uniform distribution $U[0,2\delta]$, with $\delta=0.1, 0.5, 2.5, 3.0$. The effect of the interval size saturates for approximately $\delta\ge0.5$ and $2.5$ for the left and right panel respectively. While finite size effects in the magnetization are hardly observed for $a=1.2$, fluctuations still increase with system size close to the transition. On the other hand, for $a=1.8$ no finite-size effects are observed, neither for the magnetization nor for its fluctuations: the transition is smoothed and no critical behavior is observed.}\label{mfigu2} 
\end{figure}

The initial values $x_i(t=0), i=1,\dots,N$, are independently drawn from a uniform distribution in the interval $[0,2\delta]$, for a given value of $\delta$. Obviously, by symmetry reasons, the same results would be obtained in the initial conditions were drawn from the interval $[-2\delta,0]$. In Fig.~\ref{mfigu2} we plot the average magnetization $m_0=\ensemblemean{|m|}$ as a function of  diversity $\sigma$ for different values of $a=1.2$ (left panel) and $a=1.8$ (right panel), for different system sizes $N$ and values of $\delta$.  These two values of $a$ evidence slight different behaviors: for $a=1.2$, while the sharpening finite-size effect of the magnetization is hardly seen in the plot, the fluctuations still increase with system size close to the transition (data not shown), what suggests the presence of a phase transition in the thermodynamic limit. Note that the dependence on the width of the initial condition $\delta$ is very weak and results are basically indistinguishable for $\delta\ge 0.5$. On the other hand, for $a=1.8$ there is hardly any dependence on the system size both for the magnetization and its fluctuations. The magnetization $m_0$ tends to zero smoothly with $\sigma$ and the fluctuations do not increase with system size (data not shown): the transition is smoothed and no critical behavior is present. Again, there is a dependence with the value of $\delta$ for small $\delta$ but the curves for $\delta=2.5$ and $\delta=5.0$ are indistinguishable from each other.
Summing up, for positive-definite initial conditions, the phase transition from order to disorder disappears at a value $a\approx 1.6$ (the actual value depends of the width $\delta$), such that the system shows always some degree of order for $a\gtrsim1.6$ (see the right panel of Fig.\ref{sigC}). In this sense, the ordered region is enhanced with respect to the predictions based upon the structure of the ground state.

\section{Conclusions}
\label{conclusions}
In this work we have studied the mean-field version of a Ginzburg-Landau, or $\phi^4$, model with additive quenched noise at zero-temperature. The model, that has recently been proposed in the framework of collective behavior induced by diversity \cite{tessone2006, ToralTessoneLopes2007}, is a field version of the random field Ising model well studied in the literature. As a function of diversity $\sigma$, a self-consistent theory predicts the presence of an order-disorder transition at a critical value $\sigma_c$, with mean field critical exponents that are equal than those of Landau's theory of thermal phase transitions. Numerical integrations of the dynamical equations (\ref{dxi}) are also performed for comparison. In the range of parameters where the system lacks metastable states ($a\le 1$), finite-size scaling relations show that the order parameter fluctuations diverge quadratically, rather than with $\gamma=1$ as in thermal, equilibrium, phase transitions. This suggests a violation of the fluctuation dissipation not associated to metastable effects such as lack of time translational invariance or aging \cite{young, spin1, spin2}. To explain this fact, we compute an effective Hamiltonian and argue that it cannot be readily expressed as ${\cal H}_{\rm eff}={\cal H}_{0}+Nhm$: as a consequence, the fluctuations of the order parameter cannot be straightforwardly related to the linear response, as it happens in equilibrium theory. 
In the range of parameters where metastability is likely to appear ($a>1$), stationary values typically do not reach the minimum of the Lyapunov potential, and accordingly numerical results deviate from the self-consistent theory, showing a strong dependence in the initial conditions. For a symmetrical distributed initial condition in the interval $[-\delta,+\delta]$, the ordered region is much reduced with respect to the predictions based upon the structure of the ground state of the potential. Furthermore, there is a region of values of $a$ for which a new transition from a disordered to an ordered state takes place at $\sigma_c'<\sigma_c$. In this case, diversity can not only destroy an ordered state but also induce order from a disordered metastable state. This new transition is genuinely metastable, and its location increases for increasing values of $a$, until coalescing with $\sigma_c$, where the ordered phase completely disappears. On the other hand, when the initial condition is distributed in $[0,2\delta]$, large enough values of $a$ destroy the critical behavior of the order-disorder transition and some degree of order remains at every value of the diversity $\sigma$. 

 We conclude that structural diversity can induce both the creation and annihilation of order in a nontrivial way, and deeply modify the dynamics of the diversity-free system counterpart. On the other hand, the apparent violation of the fluctuation-dissipation relation should be further investigated; at this point we can conclude that to directly relate the order parameter fluctuations to the linear response of a system can be tricky, even in the absence of metastability. This is particularly relevant in problems involving the estimation of critical exponents in nonequilibrium phase transitions.

\noindent\textbf{Acknowledgments}
The authors acknowledge financial support by the MEC (Spain) and FEDER (EU) through projects FIS2007-60327 and FIS2009-13690. NK is supported by a grant from the Govern Balear. We acknowledge useful discussions with F. Ritort, E. Koroutcheva, F.J. P\'erez-Reche and J. Fernandez.

\section*{Appendix: solutions of the cubic equation} 
\label{cardano}
We give explicit expressions for the function $\bar x(m+\eta)$ defined as the convenient real solution of the cubic equation $\alpha x+x^3=z$, where $\alpha=1-a$ and $z=m+\eta$.

In the case $\alpha\ge 0$ there is only one real solution to this equation as given by Cardano's formula

\begin{equation}
\bar x(z)=u-\alpha/(3u), \hspace{1.0cm} u=\sqrt[3]{\frac{z}{2}+\sqrt{\frac{z^2}{4}+\frac{\alpha^3}{27}}}.
\end{equation}
 
For $\alpha<0$, the same formula applies if the discriminant $\Delta\equiv 27z^2+4\alpha^3$ is positive $\Delta\ge 0$, i.e. $z\notin\left(-2(-\alpha/3)^{3/2},+2(-\alpha/3)^{3/2}\right)$. Otherwise, out of the three real solutions, the one that minimizes the local potential $v(x)=\frac{\alpha}{2} x^2+\frac{1}{4}x^4-z x$ is obtained using the trigonometric form of Cardano's formula:
\begin{equation}
\bar x(z)=2\,{\rm sign}(z)\sqrt{-\frac{\alpha}{3}}\cos\left(\frac13\arccos\sqrt{-\frac{27z^2}{4\alpha^3}}\right),
\end{equation}
where the $\arccos$ function  takes values in the principal branch $[0,\pi/2]$ of.
Note that, in every case, the function $\bar x$ is antisymmetric $\bar x(z)=-\bar x(-z)$.\\


\begin{thebibliography}{}
\bibitem{fsigma} This assertion is certainly true for a Gaussian distribution $g(\eta)$ as well as for other probability distributions, although we have not been able to give a general proof of its validity.
\end{thebibliography}
\end{document}